# Contribution of resonance tunneling of molecule to physical observables

P.M. Krassovitskiy and F.M. Pen'kov[#]

*Institute of Nuclear Physics, Almaty, Kazakhstan*

Probabilities of resonant tunneling through a potential barrier are compared for a rigid molecule and an excited molecule. It is shown that the resonance spectrum is mainly governed by the transmission resonance spectrum of the rigid molecule. Analytical expressions for the probability for the tunneling of the rigid molecule through a barrier allow resonance-spectrum-averaged observables, including quantum diffusion, to be estimated.

**Introduction.** The resonant character of the transmission probability (TP) for a bound pair incident on a potential barrier was first pointed out by Saito and Kayanuma [1] as far back as 1994. Later direct numerical solutions of the two-dimensional scattering problem allowed this effect to be confirmed for the penetration of a pair of particles bound by the oscillatory interaction through potential barriers of the Gaussian and Coulomb type [2, 3]. It was also shown that resonances occurred due to the metastable states of the pair and the barrier and the number of transmission resonances is determined by the degree of excitation of the pair. To date, the effect of resonant tunneling of a bound pair has not only been confirmed for repulsive barriers of various types but also extended to objects of more than two particles [4]. A number of works dealt with penetration through the barrier by a pair similar in spatial structure to a molecule [5, 6]. A molecular structure of the wave function for the pair was obtained by introducing interatomic interaction with repulsion at short interatomic distances. In [5] square wells separated by a square barrier were used for the interatomic potential, and in [6] the Morse potential was used for the interatomic interaction.

Treatment of a molecule as a bound pair in [5, 6] extends the field of investigations into quantum tunneling features of a bound pair to problems of quantum diffusion [7, 8] of binary molecules and becomes utterly important now that quantum diffusion of hydrogen atoms on the copper crystal surface has been directly observed [9]. In [6], for example, activation energy variation due to internal excitations of a molecule is studied.

As will be seen below, the widths of the resonances in the TP for realistic potentials are so small that estimation of their contribution to physical processes makes sense only for integrated quantities. In this work, integrated quantities of the transmission probability $W$ are taken to mean integrals like $A = \int_{E_{\min}}^{E_{\max}} f(E) W(E) dE$ with the function $f(E)$ smooth on the scale of resonance widths. This function can be the Boltzmann distribution function, i.e., $f(E) = \exp(-E/T)/T$. At $E_{min} = 0$ and $E_{max} = \infty$ the function $A$, denoted below as $F(T)$, determines diffusion in a solid (see for example [6]) or, to be more precise, that part of it which is associated with the passage of the particle through a barrier of height $V_0$. In particular, if the TP is taken to be 1 when $E \geq V_0$ and 0 when $E < V_0$, the averaging yields the classical Arrhenius equation $F(T) = \exp(-V_0/T)$.

---

[#] Corresponding author, e-mail: penkov@inp.kz

We investigated exact numerical solutions for the TP of a beryllium molecule tunneling through potential barriers on the scale of the barriers for the hydrogen atom on the copper crystal surface [9]. It turns out that when a molecule tunnels through a barrier, most of the resonance in the transmission probability are associated with the metastable states of the unexcited molecule and the repulsive barrier and allow a simple analytical description. In turn, the analytical description of the resonant tunneling allows contributions of the resonant structure to the TP-integrated quantities to be estimated.

**Tunneling of a molecule through a potential barrier.** Below we briefly formulate the problem of tunneling of a bound pair through a barrier. A more detailed formulation of the problem can be found in [1–6].

Finding the probability for a pair of identical particles with masses $m$ and coordinates $x_1$ and $x_2$ bound by the $U(x_2 - x_1)$ potential to tunnel through the potential barrier $V(x_1) + V(x_2)$ is reduced to the solution of the two-dimensional Schrödinger equation[1] [2, 3]

$$\left(-\frac{1}{2m}\Delta_1 - \frac{1}{2m}\Delta_2 + U(x_2 - x_1) + V(x_1) + V(x_2)\right)\psi = E\psi, \quad (1)$$

which can be conveniently considered in terms of the variables of the relative motion $x = x_2 - x_1$ and the center-of-mass motion $y = (x_2 + x_1)/2$

$$\left(-\frac{1}{m}\Delta_x - \frac{1}{4m}\Delta_y + U(x) + V(y + x/2) + V(y - x/2)\right)\psi = E\psi. \quad (2)$$

The potential $U(x)$ binding the atoms into the molecule determines its wave functions $\varphi_n(x)$ and spectrum $\varepsilon_n$

$$\left(-\frac{1}{m}\Delta_x + U(x)\right)\varphi_n = \varepsilon_n \varphi_n, \quad n = 1, 2, ..., N, \quad (3)$$

where $N$ is the number of the limiting excited state. We will consider a molecule in the ground state incident on the barrier with a possibility for the molecule to be in any excited state after the interaction. Then the asymptotic conditions supplementing equation (2) determine the molecule transmission and reflection amplitudes $T_{1n}$ and $R_{1n}$

$$\Psi(x, y) \xrightarrow{y \to -\infty} \varphi_1(x)\exp(ik_1 y) + \sum_{n \leq N} R_{1n}\varphi_n(x)\exp(-ik_n y),$$

$$\Psi(x, y) \xrightarrow{y \to +\infty} \sum_{n \leq N} T_{1n}\varphi_n(x)\exp(ik_n y), \qquad (4)$$

$$\Psi(x, y) \xrightarrow{x \to \pm\infty} 0.$$

---

[1] In this work the Planck and Boltzmann constants are taken to be 1, and the Kelvin (K) temperature scale is used for energy.

Here $k_n = \sqrt{4m(E - \varepsilon_n)}$ is the center-of-mass momentum of the molecule in the state with energy $\varepsilon_n$. Note that asymptotic conditions (4) determine the problem of the tunneling of the molecule through the barrier with possible excitation only to discrete states and are valid below the molecule breakup energy. In this formulation the probabilities of transmission $W_{1n} = \dfrac{k_n}{k_1}|T_{1n}|^2$ and reflection $D_{1n} = \dfrac{k_n}{k_1}|R_{1n}|^2$ with excitation to the state $n$ determine the total probabilities of transmission $W$ and reflection $D$:

$$W = \sum_{n \leq N} W_{1n}, \quad D = \sum_{n \leq N} D_{1n}. \qquad (5)$$

The condition for maintaining the probability $W + D = 1$ can be an additional accuracy criterion for numerical calculations.

To compare unambiguously the passage of the molecule and its constituent atoms through the barrier, it is desirable to use molecules with the minimum deformation of the atomic electron shells, i.e., molecules bound by the Van der Waals forces. The lightest molecule of this kind that has a series of bound states is Be$_2$ [10]. It is for these molecules that the earlier used simplified description of the interatomic interaction [5, 6] is suitable. As was already pointed out above, in [6] the interatomic interaction was described by the Morse potential, which, in addition to its simple notation, allows an analytical solution of the Schrödinger equation. In this work the potential $U(x)$ is also chosen in the form of the Morse potential

$$U(x) = U_0\left(e^{-2\rho(r - r_{eq})} - 2e^{-\rho(r - r_{eq})}\right), \qquad (6)$$

where $r = |x|$, $U_0$ is the depth of the molecular interaction well at the equilibrium point $r_{eq}$, and $\rho$ is responsible for the width of the potential well. The set of values $U_0 = 1280$ K, $r_{eq} = 2.47$ Å, and $\rho = 2.968$ Å$^{-1}$ determines the beryllium molecule with the parameters given in [10] and generates five bound states $\varepsilon_1 = -1044.88$ K, $\varepsilon_2 = -646.16$ K, $\varepsilon_3 = -342.79$ K, $\varepsilon_4 = -134.78$ K, and $\varepsilon_5 = -22.13$ K. The Be$_2$ molecule is also known to have highly excited states [11], but the five-level molecule approximation is enough for our calculations.

The repulsive barrier potential was chosen in the form of the Gaussian function

$$V(x_{1,2}) = V_0 \exp\left(-\frac{x_{1,2}^2}{2\sigma}\right) \qquad (7)$$

with the maximum value $V_0$ = 1200 K and $\sigma = 5.23 \cdot 10^{-2}$ Å$^2$. The parameter $\sigma$ determines the repulsive potential width of 1 Å at the kinetic energy of the atom in the ground state of the molecule. The quantity $V_0$ is determined by the ultimate capability of numerical calculations and has the scale of the repulsive potential for a hydrogen atom on the copper crystal surface [9].

Figure 1 shows the calculated total molecule TP $W_m$ as a function of the kinetic energy $E_k = E - \varepsilon_1$. The transmission probability has a pronounced resonant character, and at low energies it exceeds background probability values by 19 orders of magnitude. Naturally, the contribution of the resonances to all physical processes, e.g., diffusion, will be proportional to the resonance width $\Gamma$, which is $1.8 \cdot 10^{-9}$ K for the first resonance at a distance of 20 K between the neighboring resonances. This relationship between the widths of the resonances and the interval between the resonances determines the interval-average probability of $\sim 10^{-10}$, which is nine orders of magnitude higher than the lower TP envelope (background values) and ten orders of magnitude higher than the Arrhenius curve values at temperatures equal to the energy of the resonance. It is worth noting that the number of resonances during the tunneling of a molecule is considerably larger than during the tunneling of a pair bound by the oscillator interaction [2, 3]. Moreover, in the latter case resonances appear only at energies higher than the oscillator excitation energy whereas in Fig. 1 it is seen that resonances appear at energies well below 398.7 K (see the molecule spectrum above).

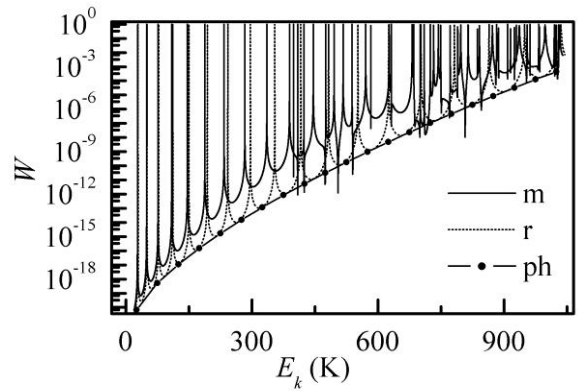

Fig. 1. Barrier transmission probabilities: "m" is $W_m$, "r" is $W_{rm}$, and "ph" is $\omega^2/4$ (see explanations in the text)

**Rigid molecule model.** The analysis shows that the above-mentioned additional resonances appear in the simple rigid molecule model when the wave function of the center of mass of the molecule corresponds to the equation

$$\left(-\frac{\hbar^2}{4m}\Delta_y + V(y+x_0/2) + V(y-x_0/2)\right)\psi(y) = E_k\psi(y) \qquad (8)$$

with the standard asymptotic conditions for the penetration of a barrier by one particle. In further calculations the fixed $x_0$ coordinate was taken to be equal to the size of the beryllium molecule, i.e., to 2.47 Å [10]. The transmission probability of the rigid molecule shown in Fig. 1 (curve "r") describes well the spectrum of the transmission resonances up to the excitation threshold of the molecule. At high energies the rigid molecule model describes only some of the resonances with a slight shift in energy. The rest of the resonances are associated with the excited states of the molecule.

In the rigid molecule model we deal with tunneling of one particle with a molecular mass through two barriers, and thus the mechanism for appearance of resonances is quite clear. Moreover, this problem allows us to obtain an expression for the two-barrier tunneling amplitude in terms of the one-barrier tunneling amplitude under rather mild limitations. Following [3], we label the particle motion regions on the left of the left barrier as 1, the region of motion between the barriers as 2, and the region of motion on the right of the right barrier by 3; i.e., transitions $1 \leftrightarrow 2$ and $2 \leftrightarrow 3$ are associated with the passage of the particle though one barrier and the transition $1 \leftrightarrow 3$ is associated with the passage of the particle through two barriers. In these terms the amplitude of the tunneling through two barriers is represented as [3]

$$T^{(13)} = \frac{T^{(12)}T^{(23)}}{1-R^{(21)}R^{(23)}}. \qquad (9)$$

Expression (9) is valid for two rectangular barriers or when quasi-classical motion conditions are fulfilled [3]. The reflection amplitude phases $R^{(21)}$ and $R^{(23)}$ differ by twice the difference of the path between the rectangular barriers or the action $S$ shifted by $\pi$ in the quasi-classical description of the motion between the barriers [3]. Since the passage through the barrier in the one-dimensional case is described in all quantum mechanics text-books, we give, without deriving them, a few relations that help simplify (9). For identical barriers both phases and moduli $T^{(12)}$ and $T^{(23)}$ are equal to each other and the amplitudes can be parameterized as $T = -i\sin\Delta\exp(i\Delta)$ and $R = \cos\Delta\exp(i\Delta)$. That is, for low transmission probabilities we have $\Delta << 1$. Denoting the transmission probability of the double-mass particle through one barrier as $\omega$, we recast (9) in the form

$$T^{(13)} = -\frac{\omega \exp(2i\Delta)}{1-(1-\omega)\exp(2i\Delta+iS)} \approx -\frac{\omega}{1-(1-\omega)\exp(iS)}. \tag{10}$$

Then the probability $W_{rm} \equiv W^{(13)} = |T^{(13)}|^2$ for tunneling through two barriers will have the form

$$W_{rm} = \frac{\omega^2}{\omega^2 - 2(1-\omega)(\cos S - 1)}. \tag{11}$$

At $S = \pm 2\pi n$ ($n = 0, 1, 2, ...$) it reaches the maximum in the transmission resonance and becomes unity, and at $S = \pi \pm 2\pi n$ it reaches the minimum value $\omega^2/4$. Figure 1 shows the "ph" curve of those minimum values. It is seen that the curve actually describes the minimum values of the TP of the rigid molecule.

Expanding $S$ near the energy of the $n$th resonance $E_n$, $S(E) \approx 2\pi n + \left.\frac{dS}{dE}\right|_{E=E_n}(E - E_n)$, we obtain the transmission amplitude and probability in this energy region in the characteristic Breit–Wigner form

$$T^{(13)} = -\frac{i\Gamma_n/2}{(E-E_n)+i\Gamma_n/2}, \quad W_{rm} = \frac{\Gamma_n^2/4}{(E-E_n)^2 + \Gamma_n^2/4}, \tag{12}$$

with the width $\Gamma_n = 2\omega \left.\frac{dE}{dS}\right|_{E=E_n}$ depending not only on the transmission probability of one particle with a molecular mass but also on the local density of the resonant states $\frac{dS}{dE}$. The equality $S(E_{n+1}) - S(E_n) = 2\pi$ allows the derivative of action to be expressed in terms of the level spacing $\left.\frac{dS}{dE}\right|_{E=E_n} \approx \frac{2\pi}{E_{n+1}-E_n}$. Naturally, this derivative can be expressed with the same accuracy in terms of the energy increments between states $n$ and $n − 1$ or in terms of the difference of energies $E_{\min,n+1}$ and $E_{\min,n}$ for the probability minima near the resonance with the number $n$. In what follows, we will therefore denote the energy increment simply by $\Delta E_n$, omitting indices of neighboring states. In this notation the width of the resonance is represented as

$$\Gamma_n \approx \frac{1}{\pi}\omega\Delta E_n. \tag{13}$$

Note that low transparency of barriers leads to small background phases $\Delta$ of the transmission amplitude of one particle passing through a barrier, which results in a characteristic shape of the resonant tunneling curve of a rigid molecule through a barrier. It is evident from Fig. 1 that this curve has no dips typical of resonant scattering in the presence of the background phase while resonances occurring behind the excitation threshold and corresponding to the *x*-directed vibrations of the molecule [3] show all typical signs of resonant scattering.

**Integrated contributions of resonances.** For calculating integrated quantities like $A = \int_{E_{\min}}^{E_{\max}} f(E) W_{rm}(E) dE$ with the function $f(E)$ smooth on the scale of resonance widths and with quite a lot of resonances within the limits of integration it is convenient to break up the integral $A$ into a sum of integrals near each resonance

$$A = \int_{E_{\min}}^{E_{\max}} f(E) W_{rm}(E) dE = \sum_n \int_{E_{\min,n}}^{E_{\min,n+1}} f(E) W_{rm}(E) dE = \sum_n A_n. \tag{14}$$

Considering small variations in the functions $f(E)$ and $\omega(E)$ on the interval $[E_{\min,n}, E_{\min,n+1}]$, integrals $A_n$ can be calculated using (11) after the integration variable $E$ is replaced with $S$

$$A_n \approx f(E_n) \frac{dE}{dS}\bigg|_{E=E_n} \int_{-\pi}^{\pi} \frac{\omega^2}{\omega^2 - 2(1-\omega)(\cos S - 1)} dS = f(E_n) \Delta E_n \frac{\omega(E_n)}{2 - \omega(E_n)}. \tag{15}$$

If there are a lot of resonances within the integration limits, the sum can be replaced with an integral

$$\sum_n f(E_n) \Delta E_n \frac{\omega(E_n)}{2 - \omega(E_n)} \approx \int_{E_{\min}}^{E_{\max}} f(E) \frac{\omega}{2 - \omega} dE. \tag{16}$$

Considering (14)–(16), we obtain an approximate equality

$$A = \int_{E_{\min}}^{E_{\max}} f(E) W_{rm}(E) dE \approx \int_{E_{\min}}^{E_{\max}} f(E) \frac{\omega}{2 - \omega} dE, \tag{17}$$

for expressing averages of $W_{rm}$ in terms of the integral of the TP of a molecular-mass particle passing through one barrier. The more resonances are within the integration limits, the more accurately equality (17) holds. It is easy to see that the first terms of the expansion $\frac{\omega}{2-\omega} = \frac{\omega}{2} + \frac{\omega^2}{4} + ...$ correspond to the integral of the resonance in the form (12) and to the contribution of the above-mentioned background values of the TP. Note that the function $\frac{\omega}{2-\omega}$ tends to 1 when $\omega$ tends to 1 and can be replaced with the asymptotic value at the infinite limits of integration.

Fulfillment of (17) is demonstrated in Fig. 2 by the $F(T)$ curves of the TP averaging over the Boltzmann distribution for three probabilities

$$F_m(T) = \beta \int_0^{E_{max}} \exp(-\beta E) W_m(E) dE + \exp(-\beta E_{max}),$$

$$F_{rm}(T) = \beta \int_0^{E_{max}} \exp(-\beta E) W_{rm}(E) dE + \exp(-\beta E_{max}), \quad (18)$$

$$F_p(T) = \beta \int_0^{E_{max}} \exp(-\beta E) \frac{\omega(E)}{2-\omega(E)} dE + \exp(-\beta E_{max}),$$

where $\beta = 1/T$. Asymptotics on the right-hand side is added because the energy range in the numerical TP calculations is limited. The upper limit of integration $E_{max}$ for the curves in Fig. 2 was 2300 K. At these energies the TP for both the rigid molecule and the double-mass particle is 1 to a high accuracy while the TP of the molecule shows nonmonotonic behavior with the average probability value 0.8. As was mentioned above, at energies above the molecule breakup threshold 1044.88 K asymptotic conditions (4) do not ensure maintaining the probability $W + D = 1$ because of neglected breakup modes. At the energy of 2300 K the probability of being in the discrete spectrum channels is 0.996, i.e., the breakup channel only slightly affects the integrated curves in Fig. 2.

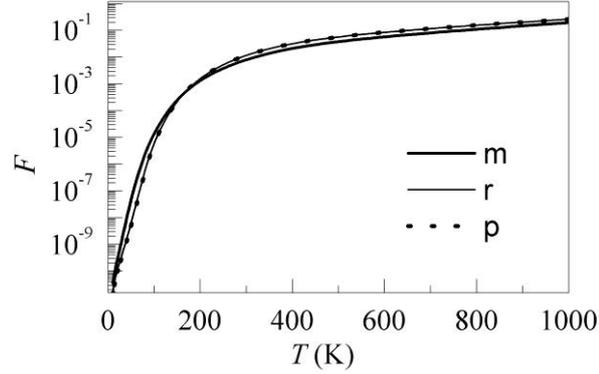

Fig. 2. Functions $F(T)$ in different models: "m" is $F_m(T)$, "r" is $F_{rm}(T)$, and "p" is $F_p(T)$ (see explanations in the text)

It is seen in Fig. 2 that the $F(T)$ values obtained from averaging the TP in the rigid molecule model (curve "r") and the TP of one molecular-mass particle (curve "r") almost coincide and confirm equality (17). Both curves are close to the curve obtained by the Boltzmann averaging of the molecule TP (curve "m"). Actually, we can say that the averaging of a very complicated resonant transmission curve of a molecule can be replaced by the averaging over the monotonic transmission probability of one particle passing through one barrier.

**Conclusions.** Comparisons were made with the simplest version of the rigid molecule model with the fixed coordinates $x = x_0$ (see equation (8)). A more complex version can be obtained by using ground-state-average repulsive potentials $\int_{-\infty}^{\infty} \left( V(y + x/2) + V(y - x/2) \right) \varphi_1^2(x) dx$ in (8). This approximation was also tested. It was found that the curves from Figs. 1 and 2 in this approximation simply overlap the curves of the model used. This implies that the difference between the models is insignificant, and we chose the simplest model for describing the results.

Similarity of the TP averaging curves for the molecule and the rigid molecule allows the conclusion that the requirements on the choice of a molecule for studying the resonant tunneling through the barrier described at the beginning of this paper are rather steep. Since the main contributions to the averages come from the resonances of the ground-state molecule TP, molecules with an arbitrary complex mechanism for molecular spectrum formation can be considered. Thus, the conclusions of this work are also applicable to, say, a hydrogen or helium molecule tunneling through a barrier. The fact that average TPs of a rigid molecule are described in terms of average TPs of one molecular-mass particle passing through one barrier allows easily estimating the contribution from resonant TPs to physical phenomena, e.g., diffusion in solids. For example, diffusion coefficients in solids at low temperatures for a hydrogen molecule in the state with the electron configurations close to the electron states of isolated hydrogen atoms should be half as large as diffusion coefficients of deuterium atoms.

The authors are grateful to S.I. Vinitsky and A.A. Gusev for the possibility of testing the calculations with alternative software. The work was supported by the Ministry of Education and Science of the Republic of Kazakhstan, grant 0602/GF.